\documentclass[a4paper,11pt,notitlepage]{article}
\usepackage{amsfonts,amssymb,amsmath,amsthm,hyperref,colonequals}
\usepackage{colonequals,simplewick}
\usepackage[colorinlistoftodos]{todonotes}
\usepackage{caption}
\usepackage{cite}
\usepackage[parsep]{collref}
\usepackage{tabularx}
\usepackage{graphicx}
\usepackage{shuffle}
\usepackage[nice]{nicefrac} 
\usepackage{url}
\theoremstyle{definition}
\newcommand{\beqa}{\begin{eqnarray}}
\newcommand{\eeqa}{\end{eqnarray}}
\newcommand{\beq}{\begin{equation}}
\newcommand{\eeq}{\end{equation}}

\newcommand{\calO}{\mathcal{O}}

\newcommand{\calN}{\mathcal{N}}

\DeclareMathOperator{\Tr}{Tr}

\newcommand{\vac}[1]{\ensuremath{\left< \, #1\, \right>}}

\newcommand{\dd}{\mathrm{d}}
\newcommand{\AAA}{\mathcal{A}}

\begin{document}

\thispagestyle{empty}
\setcounter{page}{0}
\begin{flushright}\footnotesize
\texttt{HU-Mathematik-2014-30}\\
\texttt{HU-EP-14/41}\\
\texttt{CERN-PH-TH-2014-200}\\
\vspace{0.5cm}
\end{flushright}
\setcounter{footnote}{0}

\begin{center}
{\huge{
\textbf{A Twistorial Approach \vspace{0.2cm}\\ to Integrability in $\calN=4$ SYM}
}}
\vspace{15mm}

{\sc 
Laura Koster, Vladimir Mitev, Matthias Staudacher }\\[5mm]

{\it Institut f\"ur Mathematik, Institut f\"ur Physik und IRIS Adlershof,\\
Humboldt-Universit\"at zu Berlin,\\
Zum Gro{\ss}en Windkanal 6, 12489 Berlin, Germany
}\\[5mm]

\texttt{laurakoster@physik.hu-berlin.de}\\
\texttt{mitev@math.hu-berlin.de}\\
\texttt{matthias@mathematik.hu-berlin.de}\\[25mm]

\textbf{Abstract}\\[2mm]
\end{center}
While the achievements in the study of  $\calN=4$ Super Yang-Mills through the application of integrability are impressive, the precise origins of the exact solvability remain shrouded in mystery. In this note, we propose that viewing the problem through the lens of twistor theory should help to clarify the reasons for integrability. We illustrate the power of this approach by rederiving the model's one-loop spin chain dilatation operator in the SO$(6)$ sector.
 

\newpage
\setcounter{page}{1}


\addtolength{\baselineskip}{5pt}

The simplest non-abelian gauge theory in four space-time dimensions is certainly the $\mathcal{N}=4$ Super Yang-Mills (SYM) model. Its rather intricate action is given, before gauge fixing, by
\beq\label{N=4action}
S=\frac{1}{g^2_{YM}}\int \dd^4x\: \Tr \left(-\frac{1}{4}F_{\mu\nu}F^{\mu\nu} +\frac{1}{2} \mathcal{D}_\mu\phi_{i}\mathcal{D}^\mu\phi^{i}-\frac{1}{8}[\phi_{i},\phi_{j}][\phi^{i},\phi^{j}]+\cdots \right),
\eeq
where we did not spell out the fermionic part.
It is conjectured that this model becomes quantum integrable at arbitrary finite coupling $\lambda=N g_{YM}^2$ in the planar $N \rightarrow \infty$ limit. The source of integrability, a hidden infinite-dimensional symmetry, is quite unclear at finite $\lambda$. In particular, the action \eqref{N=4action} gives no immediate indication for its presence.  Actually, already the effect of the finite-dimensional superconformal symmetry algebra $\mathfrak{psu}(2,2|4)$ on the fields in the action is quite involved. One is tempted to think that a more natural form of the action should be such that the symmetry algebra acts linearly, which is the case in the twistorial formulation of $\mathcal{N}=4$ SYM. Another indication that the twistor formalism might be more suitable for studying integrability in $\mathcal{N}=4$ SYM is the fact that its action's leading order component is equivalent to the self-dual Yang-Mills action in space-time which has been shown to be classically integrable, see \cite{Mason:1991rf} and references therein. Thus, we are encouraged to investigate the appearance of integrability in this approach. Motivated by the fact that the quantum integrability of $\mathcal{N}=4$ SYM is most easily detected in the derivation of the one-loop dilatation operator in the SO$(6)$ sector \cite{Minahan:2002ve, Minahan:2010js}, we revisit this computation from a twistorial point of view. 

Before we begin our illustrative computation, we shall give a brief introduction to the twistor space formalism. We mostly follow the notation and conventions used in \cite{Bullimore:2013jma, Adamo:2013cra}, where further details about the formalism, for example the construction of the twistor action, can be found.
One starts with the complexification of conformally compactified Minkowski space $\mathbb{CM}^{\#}$, which admits a description as a complex quadric surface in $\mathbb{CP}^5$. This can be seen by introducing six homogeneous coordinates on $\mathbb{CP}^5$ transforming in the anti\-symmetric tensor representation of the complexified conformal group SL$(4,\mathbb{C})$, 
\begin{equation}
X^{IJ}=-X^{JI}\,.
\end{equation}
We can then identify $\mathbb{CM}^{\#}$  with the quadric on $\mathbb{CP}^5$ that is given by the equation
\begin{equation}
X\cdot X\colonequals \frac{1}{2} \epsilon_{IJKL} X^{IJ}X^{KL}=0\,.
\end{equation}
We now introduce the bosonic twistor space as the complex projective space $\mathbb{CP}^3$ whose elements $Z_I$, called twis\-tors,  transform in the fundamental representation of SL$(4,\mathbb{C})$. This allows us to identify  the points in $\mathbb{CM}^{\#}$ as lines in twistor space via the \textit{incidence relation}
\begin{equation}\label{eq:incid}
X^{IJ}Z_J=0\,.
\end{equation}
Given the line $\{Z\in \mathbb{CP}^3: X^{IJ}Z_J=0\}$ in twistor space, one can reconstruct the point $X^{IJ}$ in $\mathbb{CM}^{\#}$ from two twistors $Z_A^I$ and $Z_B^J$ satisfying \eqref{eq:incid} via $X^{IJ}=Z_A^{[I}Z_B^{J]}$. By abuse of notation $X$ will refer both to a point in $\mathbb{CM}^{\#}$ and to the associated line in twistor space.
It is convenient to choose local affine coordinates on $\mathbb{CP}^5$, or equivalently, a point at infinity $I^{IJ}$ and to divide $\mathbb{CM}^{\#} $ by the so-called `light cone at infinity', defining
\begin{equation}
\mathbb{CM}\colonequals\mathbb{CM}^{\#}/\{ X\in\mathbb{CM}^{\#} \:|\: X\cdot I =0\} \,.
\end{equation}
The point at infinity then can be used to construct a metric on $\mathbb{CM}$ via
\begin{equation}
\label{eq:defmatric}
g(X,Y)=\frac{X\cdot Y}{(I\cdot X)(I\cdot Y)}\,.
\end{equation}
The fundamental SL$(4,\mathbb{C})$ indices are decomposed into spinor SL$(2,\mathbb{C})\times \text{SL}(2,\mathbb{C})$ indices $\alpha=1,2$ and $\dot\alpha=1,2$, 
\begin{equation}
Z^I=(\lambda_{\alpha},\mu^{\dot{\alpha}})\,.
\end{equation} 
This point at infinity  may be chosen as
\begin{equation}
I_{IJ}=
\left( \begin{array}{cc}
\epsilon^{\alpha\beta}&0\\
0&0\end{array}\right),
\qquad I^{IJ}=
\left( \begin{array}{cc}
0&0\\
0&\epsilon^{\dot{\alpha}\dot{\beta}}\end{array}\right),
\eeq
allowing us to introduce coordinates $x^{\alpha\dot{\alpha}}$ on $\mathbb{CM}$ as
\begin{equation} X^{IJ}=
\left( \begin{array}{cc}
\epsilon_{\alpha\beta}&-ix_{\alpha}^{\phantom{\alpha}\dot{\beta}}\\
ix^{\phantom{\beta}\dot{\alpha}}_{\beta}&-\frac{1}{2}x^2\epsilon^{\dot{\alpha}\dot{\beta}}\end{array}\right)\quad  \text{ or } \quad X_{IJ}=
\left( \begin{array}{cc}
-\frac{1}{2}x^2\epsilon^{\alpha\beta}&ix^{\alpha}_{\phantom{\alpha}\dot{\beta}}\\
-ix_{\phantom{\beta}\dot{\alpha}}^{\beta}& \epsilon_{\dot{\alpha}\dot{\beta}}\end{array}\right),
\end{equation}
where 
\begin{equation}x^{\alpha\dot{\alpha}}=
\frac{1}{\sqrt{2}}\left( \begin{array}{cc}
x_0+x_3&x_1-ix_2\\
x_1+ix_2&x_0-x_3\end{array}\right)^{\alpha\dot{\alpha}}\,.\end{equation}
 In these coordinates, the solutions to the incidence relation \eqref{eq:incid} take the form
\beq\label{eq:solinc}
Z^I=(\lambda_{\alpha},ix^{\alpha\dot{\alpha}}\lambda_{\alpha})\,.
\eeq
Super Minkowski space can be defined by introducing the eight additional Gra\ss mann coordinates $\theta^{\alpha a}$. Similarly, twistor space can be extended to supertwistor space $\mathbb{CP}^{3|4}$. The supertwistors $\mathcal{Z}\in\mathbb{CP}^{3|4}$ lying on the line $X$ that now corresponds to the super space-time point $(x,\theta)$ are given by
\beq
\mathcal{Z}^I=(\lambda_{\alpha},\mu^{\dot{\alpha}},\chi^a)=(\lambda_{\alpha},ix^{\alpha\dot{\alpha}}\lambda_{\alpha},\theta^{\alpha a}\lambda_{\alpha})\,,
\eeq where the index $a$ runs from $1$ to $4$.
In the rest of this paper we will continue to use calligraphic $\mathcal{Z}$ to denote the supertwistor and ordinary $Z$ to denote its bosonic part. We define a supertwistor field $\AAA(\mathcal{Z})$ as a $(0,1)$-form on twistor space which can be expanded in the Gra{\ss}mann variables $\chi$ as \cite{Nair:1988bq}
\begin{equation}
\label{eq:expansionAAA}
\AAA(\mathcal{Z})=a +\chi^a\tilde{\psi}_a+\frac{1}{2}\chi^a\chi^b\phi_{ab}+\frac{1}{3!}\chi^a\chi^b\chi^c\psi^d\epsilon_{abcd}+\chi^1\chi^2\chi^3\chi^4 g\,,
\end{equation} where the fields $a$, $\tilde{\psi}$, $\phi$, $\psi$ and $g$ depend only on the bosonic twistor $Z$ (and $\bar{Z}$)\,. We note that the scalars $\phi_{ab}(x)$ of \eqref{eq:expansionAAA} in SU$(4)$ index notation are related to the SO$(6)$ index ones $\phi_j$ of \eqref{N=4action} via $\phi_{12}=\phi_1+i\phi_2$, et cetera.
Fields on twistor space can be related to fields on space-time via the Penrose transform. More precisely,  a Dolbeault cohomology class of fields on twistor space that are homogeneous of degree $n$ is isomorphic to a solution of the zero-rest-mass field equations of helicity $h=(n+2)/2$. For example, for a scalar field in space-time this isomorphism is realized as
\begin{equation}
\label{eq:singlephifield}
\phi(x)=\frac{1}{2\pi i}\int_{X} \langle \lambda\dd\lambda\rangle \:h^{-1}(x,\lambda)\phi (\lambda_{\alpha},ix^{\alpha\dot{\alpha}}\lambda_{\alpha})h(x,\lambda)\,,
\end{equation}
where $\lambda=(\lambda_1,\lambda_2)$ parametrizes the projective line $X\simeq \mathbb{CP}^1$ in twistor space corresponding to the space-time point $x$ and $\langle \lambda \lambda' \rangle \colonequals \lambda_{\alpha}\lambda'_{\beta}\epsilon^{\alpha \beta}$ with $\epsilon^{12}=1$. The function $h(x,\lambda)$ is a holomorphic frame that depends smoothly on the line $X$, see for instance (2.32\textit{ff.})  of \cite{Bullimore:2013jma}.
One can define an action for the supertwistor field $\AAA(\mathcal{Z})$ as the sum of a holomorphic Chern-Simons action, introduced in \cite{Witten:2003nn},
\begin{equation}
S_1=\frac{i}{2\pi}\int D^{3|4}\mathcal{Z} \:\:\Tr\,(\AAA\wedge\bar{\partial}\AAA+\frac{2}{3}\AAA\wedge\AAA\wedge\AAA)\,, 
\end{equation} 
where $ D^{3|4}\mathcal{Z}\colonequals  \frac{1}{4!}\epsilon_{IJKL}Z^I\dd Z^J\dd Z^K\dd Z^L\dd^4\chi$, 
and an `interaction' part\footnote{The normalization of the action is determined by deriving the usual space-time action from $S_1+S_2$.}
\begin{equation}
S_2=-\frac{g_{YM}^2}{4}\int \dd^4 z\:\dd^8\theta \:\log\det \,(\bar{\partial}+\AAA)_{(z,\theta)}\,,
\end{equation} which first appeared in \cite{Boels:2006ir}. Remark that interactions in twistor space are localized on the \textit{line} corresponding to the super space-time point $(z,\theta)$. By expanding the logarithm in $S_2$ we obtain 
\begin{multline}
\label{eq:interactionexpanded}
S_2=-\frac{g_{YM}^2}{4}\int \dd^4 z\:\dd^8\theta \:\Bigg[\:\Tr \log \,(\bar{\partial}_{(z,\theta)})\\+\sum_{n=1}^{\infty}\int_{\mathbb{C}^n}\frac{1}{n}\left(\frac{-1}{2\pi i}\right)^n \frac{\dd\rho_1\dd\rho_2\cdots\dd\rho_n}{\rho_{1n}\rho_{n(n-1)}\cdots \rho_{21}}
\Tr\big(\AAA(\rho_1)\cdots\AAA(\rho_n)\big)\Bigg],
\end{multline}
where $\rho_{ij}\colonequals\rho_i-\rho_j$. From now on, we drop the infinite constant $\Tr \log (\bar{\partial}_{(z,\theta)})$ and are left with an infinite sum of  $n$-vertices. Those $n$-vertices in the second line of \eqref{eq:interactionexpanded} are given as integrals over the supertwistor line corresponding to the super space-time point $(z,\theta)$. We define  this line by two supertwistors $\mathcal{Z}_C$ and $\mathcal{Z}_D$ and parametrize the $i$-th twistor on it using affine coordinates $\rho_i\in \mathbb{C}$ by $\mathcal{Z}(\rho_i)=\mathcal{Z}_C+\rho_i \mathcal{Z}_D$, so that $\AAA(\rho_i)=\AAA(\mathcal{Z}(\rho_i))$.

 Note that the twistor action has a rather compact form compared to the usual $\mathcal{N}=4$ SYM action, a feature that we will exploit in our calculation of the one-loop dilatation operator. 
It was shown in \cite{Boels:2006ir} that the total action $S_1+S_2$ reduces in a partial gauge, termed harmonic gauge, to the standard action \eqref{N=4action} of $\mathcal{N}=4$ SYM, with no space-time gauge fixing imposed. In the current paper, we will not use this harmonic gauge, but rather the axial gauge that was first introduced in \cite{Cachazo:2004kj} and was defined with respect to a `twistor at infinity' $Z_*$. The gauge condition reads
 \begin{equation}
\overline{Z_* \cdot \frac{\partial}{\partial Z}}\:\lrcorner\:\AAA=0\, ,
\end{equation} 
with $\lrcorner$ denoting the interior product. In this gauge, the super field $\AAA$ has only two independent components and therefore the cubic term in the holomorphic Chern-Simons action $S_1$ vanishes. Following the reasoning in \cite{Adamo:2011cb}, the propagator of $\AAA$ is derived from the quadratic part of $S_1$, while the quadratic part of $S_2$ is treated as a two-point vertex. Thus the propagator is the inverse of $\frac{i}{2\pi}\bar{\partial}$\, :
\begin{equation}
\langle \AAA(\mathcal{Z}_1)\AAA(\mathcal{Z}_2)\rangle =-\bar{\delta}^{2|4}(\mathcal{Z}_1,*,\mathcal{Z}_2)\colonequals -\int_{\mathbb{C}^2}\frac{\dd s\dd t}{s t }\bar{\delta}^{4|4}(\mathcal{Z}_1+s\mathcal{Z}_*+t\mathcal{Z}_2)\,,
\end{equation}
where the $\delta$ ``function'' on the right hand side is actually the $(0,4)$-form on twistor space given by
\beq
\bar{\delta}^{4|4}(\mathcal{Z})\colonequals\bigwedge_{I=1}^4\delta(Z^I)d\bar{Z}^I\prod_{a=1}^4\chi^a\, .
\eeq

We now have all the necessary tools to compute the dilatation operator in the SO$(6)$ sector to one loop. We define the gauge invariant operators
\begin{align}
\label{eq:defoperatorO}
\mathcal{O}_{\textbf{I}}(x)\colonequals&\, \Tr \,(\phi_{a_1b_1}\cdots \phi_{a_Lb_L})(x)\\=&\int_{X^L}\left(\prod_{l=1}^L \frac{\langle \lambda_l\dd\lambda_l\rangle}{2\pi i }\frac{\partial^2}{\partial\chi_l^{a_l}\partial \chi_l^{b_l}}\right)  
\Tr\big(\AAA(\mathcal{Z}_1) U(\mathcal{Z}_1,\mathcal{Z}_2)\cdots \AAA(\mathcal{Z}_L)U(\mathcal{Z}_L,\mathcal{Z}_1)\big)\,,\nonumber
\end{align}
where $\mathcal{Z}_l=(\lambda_{l \alpha},\mu_l^{\dot{ \alpha}}, \chi_l^{a})=(\lambda_{l \alpha}, ix^{\alpha\dot{\alpha}}\lambda_{l \alpha}, \theta_l^{\alpha a}\lambda_{l \alpha})$, and all the $\wedge$-symbols between the forms are omitted. Note that the Gra{\ss}mann variables $\theta_l^{\alpha a}$ span an independent Gra{\ss}mann algebra at each site. The parallel propagators $U(\mathcal{Z}_l,\mathcal{Z}_{l+1})$ are given by concatenating two frames as $H(x,\theta,\lambda_l)H^{-1}(x,\theta,\lambda_{l+1})$, where the $H$ are supersymmetric generalizations of the frames $h$ of \eqref{eq:singlephifield}, see (2.54\textit{ff}) of \cite{Bullimore:2013jma}. These parallel propagators $U$ can be expressed as path-ordered exponentials of $\AAA$ and serve to guarantee the gauge invariance of the operator $\calO_{\textbf{I}}$, see (4.10) of \cite{Adamo:2013cra}. At tree level, a straightforward counting of Gra{\ss}mann variables shows that the parallel propagators $U(\mathcal{Z}_l,\mathcal{Z}_{l+1})$ do not contribute. However, at one loop the parallel propagators $U(\mathcal{Z}_l,\mathcal{Z}_{l+1})$ in \eqref{eq:defoperatorO} require a more careful treatment and can in principle contribute to the correlation functions that we wish to compute. A detailed study and interpretation of the parallel propagators will be one of  the subjects of a forthcoming publication \cite{KMSW}, where it will be shown in detail how their contributions cancel at one loop. Thus, we shall ignore them here, which allows us to only consider the simplified expression
\begin{align}
&\mathcal{O}_{\bf{I}}(x)=\int_{X^L}\left( \prod_{l=1}^L\frac{\langle \lambda_l\dd\lambda_l\rangle}{2\pi i}\right)\,    \Tr \Big(\frac{\partial^2 \AAA(\mathcal{Z}_1)}{\partial\chi_1^{a_1}\partial \chi_1^{b_1}}\cdots \frac{\partial^2\AAA(\mathcal{Z}_L)}{\partial\chi_L^{a_L}\partial \chi_L^{b_L}}\Big).
\end{align} 
We now compute the two-point correlation functions of the fields $\calO_{\textbf{I}}$. First, we look at the tree-level correlator $\vac{\mathcal{O}_{\textbf{I}}(x_1)\mathcal{O}_{\textbf{I}'}(x_2)}$. Here we parametrized the lines corresponding to the space-time points $X_k$ (with $k=1,2$) us\-ing two fixed  su\-per twistors $\mathcal{Z}_{A_k}=(A_{k \alpha},ix_k^{\alpha\dot{\alpha}} A_{k \alpha},\chi_{A_k}^a)$ and $\mathcal{Z}_{B_k}=(B_{k \alpha},ix_k^{\alpha\dot{\alpha}} B_{k \alpha},\chi_{B_k}^a)$ on each line, via the equation $\mathcal{Z}_k(s)=\mathcal{Z}_{A_k}+s\mathcal{Z}_{B_k}$. Furthermore, in what follows $(A_1 B_1 A_2B_2)$ is defined as the determinant of the $4\times 4$ matrix that has the four bosonic twistors $Z_{A_1}$, $Z_{B_1}$, $Z_{A_2}$, and $Z_{B_2}$ as its columns. The tree-level correlator $\vac{\mathcal{O}_{\textbf{I}}(x_1)\mathcal{O}_{\textbf{I}'}(x_2)}$ reduces to simple Wick contractions of the form (we suppress the color indices)
\begin{align}
\label{eq:treelevelpropagator}
\langle \phi_{ab}(x_1)\phi_{cd}(x_2)\rangle &=\int_{X_1}\frac{\langle \lambda_1\dd\lambda_1\rangle}{2 \pi i } \int_{X_2}\frac{\langle \lambda_2\dd\lambda_2\rangle}{2\pi i}\vac{\frac{\partial^2 \AAA(\mathcal{Z}_1)}{\partial\chi_1^a\partial \chi_1^b}\frac{\partial^2 \AAA(\mathcal{Z}_2)}{\partial \chi_2^c\partial \chi_2^d}}
\nonumber\\
&=\int_{\mathbb{C}}\frac{\langle A_1 B_1\rangle \dd s}{(2\pi) s} \int_{\mathbb{C}}\frac{\langle A_2 B_2\rangle \dd t}{(2\pi) t}\frac{\partial^2}{\partial\chi_{A_1}^a\partial \chi_{B_1}^b}\frac{\partial^2}{\partial\chi_{A_2}^c\partial \chi_{B_2}^d}\bar{\delta}^{2|4}(\mathcal{Z}_1(s),*,\mathcal{Z}_2(t))\nonumber\\
& = \frac{\epsilon_{abcd}}{(2\pi)^2}  \frac{\langle A_1 B_1\rangle \langle A_2 B_2\rangle}{(A_1B_1A_2B_2)}=\frac{1}{(2\pi)^2}\frac{2\epsilon_{abcd}}{|x_1-x_2|^2}\, ,
\end{align}
where the last equality follows by inserting $X^{IJ}_k=Z_{A_k}^{[I}Z_{B_k}^{J]}$ into \eqref{eq:defmatric}.
Thus, using the same combinatorics as in space-time, we can find the desired tree-level correlation function.

As in space-time, computing the two-point correlation functions of the operators $\calO$ at  one loop reduces to the calculation of the subcorrelator
\begin{equation}\label{subcor}
\langle\big( \phi_{ab}\phi_{cd}\big)^i_{\phantom{i}j}(x_1) \big(\phi_{a'b'}\phi_{c'd'}\big)^k_{\phantom{k}l}(x_2) \rangle_{1-\mathrm{loop}}\, ,
\end{equation} where $a,b,c,d,a',b',c',d'$ are SU$(4)$ flavor indices and $i,j,k,l$ are color indices. 
 Even though there is an infinite number of vertices in \eqref{eq:interactionexpanded}, only the four-vertex has a contribution at one loop. 
 \begin{figure}[ht]
 \centering
  \includegraphics[height=4cm]{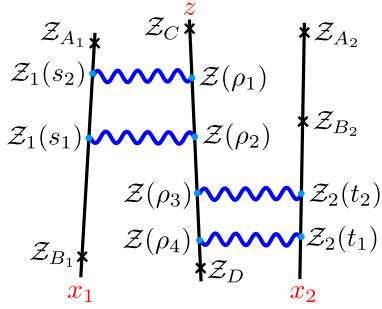}
  \caption{\it The figure shows the only twistor diagram that contributes to the one-loop dilatation operator. The 
  $Z$'s are points on the bosonic part of supertwistor-space, and the twistor lines are labeled by the space-time points $x_1, z, x_2$. Each line is defined by two fixed twistors, whose positions are marked by crosses. Twistor propagators are represented by wavy lines.}
  \label{fig:1loopdiagram}
\end{figure}
This is due to the fact that the propagator $\contraction{}{\phi}{}{\AAA}\phi\AAA$
is of second order in the Gra{\ss}mann variables, while eight Gra{\ss}mann numbers are needed for a non-trivial integration over  $\dd^8\theta$. 
This means that, contrary to the situation in space-time, see e.g. \cite{Minahan:2002ve, Minahan:2010js}, only one type of planar diagram will be relevant at one loop. It is illustrated in figure~\ref{fig:1loopdiagram}. 
The subcorrelator \eqref{subcor} thus equals  (we suppress the color indices)
\begin{align}\label{dil}
&\frac{1}{(2\pi)^8}  \frac{g^2_{YM}N^2}{4}\int\dd^{4}z\:\dd^8\theta\int\frac{\dd s_1}{s_1} \int\frac{\dd s_2}{s_2} \int\frac{\dd t_1}{t_1} \int\frac{\dd t_2}{t_2}\int \frac{\dd\rho_4\dd\rho_3\dd\rho_2\dd\rho_1}{\rho_{14}\rho_{43}\rho_{32}\rho_{21}}\\
&\times (\langle A_1B_1\rangle \langle CD\rangle)^2 \frac{\partial^2}{\partial\chi_1^a\partial\chi_1^b}\Big(\bar{\delta}^{2|4}(\mathcal{Z}_1(s_1),*,\mathcal{Z}(\rho_2))\Big)\frac{\partial^2}{\partial\chi_1^c\partial \chi_1^d}\Big(\bar{\delta}^{2|4}(\mathcal{Z}_1(s_2),*,\mathcal{Z}(\rho_1))\Big)\notag\\
&\times (\langle A_2B_2\rangle \langle CD\rangle)^2\frac{\partial^2}{\partial\chi_2^{a'}\partial\chi_2^{b'}}\Big(\bar{\delta}^{2|4}(\mathcal{Z}_2(t_1),*,\mathcal{Z}(\rho_4))\Big)\frac{\partial^2}{\partial\chi_2^{c'}\partial\chi_2^{d'}}\Big(\bar{\delta}^{2|4}(\mathcal{Z}_2(t_2),*,\mathcal{Z}(\rho_3))\Big).\notag
\end{align}
Let us make a few observations. 
We first remark that \eqref{dil} must be SO$(6)$ invariant, therefore we can write it  (in SU$(4)$ notation) as the linear combination
\begin{equation}
 \textsf{A}\epsilon_{abcd}\epsilon_{a'b'c'd'}+\textsf{B}\epsilon_{aba'b'}\epsilon_{c'd'cd}+\textsf{C}\epsilon_{abc'd'}\epsilon_{a'b'cd}\,,
\end{equation} so that the problem translates to determining\footnote{One way of obtaining the coefficients $ \textsf{A}$, $ \textsf{B}$ and $ \textsf{C}$ is to set the flavor indices  to specific values (for example, $(abcd)=(1213)$ and $(a'b'c'd')= (3424)$ will yield $- \textsf{B}$).} $ \textsf{A}$, $ \textsf{B}$ and $ \textsf{C}$. 
The second observation is that there are eight bosonic delta functions and also precisely eight fiber integrations. Thus, the integrations over the fiber variables $s_i$,  $t_j$,   $\rho_k$, amount to simply evaluating on the support of the delta functions. For example, $\rho_1$ will be fixed by the five bosonic twistors $Z_{A_1}$, $Z_{B_1}$,  $Z_{*}$, $Z_{C}$ and $Z_D$ via Cramer's rule to $\rho_1=\tfrac{(C*A_1B_1)}{(D*A_1B_1)}$. 
This implies not only that all fiber integrations are essentially trivial but also that $\rho_1$ and $\rho_2$ (resp.\ $\rho_3$ and $\rho_4$) are forced to be equal. 
Therefore, after performing the fiber integrals, $1/(\rho_2-\rho_1)(\rho_4-\rho_3)$ is naively $1/0$\footnote{We stress that whenever we refer to $1/0$-like divergences we are referring to the unphysical divergences that arise from evaluating $1/(\rho_1-\rho_2)(\rho_3-\rho_4)$ on the support of the delta functions on the fibers and \textit{not} to the UV divergences that arises from $z\rightarrow x_1$ or $z\rightarrow x_2$.}. 
However, if one first performs the Gra{\ss}mann integrals, one finds that $ \textsf{B}$ and $ \textsf{C}$ contain a factor of $(\rho_2-\rho_1)(\rho_4-\rho_3)$ that cancels the identical factor in the denominator and hence no $1/0$-like divergences occur after subsequently evaluating the bosonic delta functions. The term $ \textsf{A}$ on the other hand does contain $1/0$-like divergences, even after first performing the Gra{\ss}mann integrations and therefore requires regularization.
Since the divergences arise from the fact that the external scalar fields are pairwise located at the same point in Minkowski space, we can resolve these divergences by using a point-splitting procedure (which translates to a \textit{line-splitting} in twistor space, see figure~\ref{fig:splitting}), for both $x_1$ and $x_2$. 
\begin{figure}[ht]
 \centering
  \includegraphics[height=3cm]{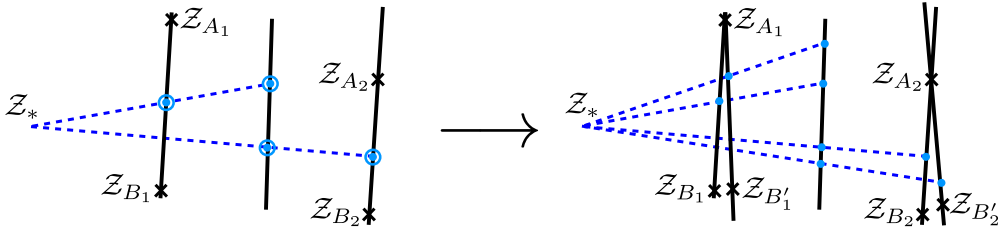}
  \caption{\it The figure illustrates our regularization procedure in twistor space. Separating the points in space-time amounts to splitting the lines in twistor space. This regularizes the Green's functions on the integration line. The dashed lines are lines in twistor space going through the reference twistor $Z_*$ and the two twistors connected by a propagator.}
  \label{fig:splitting}
\end{figure}
One needs to be careful to perform this splitting without breaking the cyclic symmetry of the operator. In our calculation this amounts to making the following replacements
\begin{align}
\bar{\delta}(\rho_{1}-a)\bar{\delta}(\rho_{2}-a)&\rightarrow \frac{1}{2}\left[\bar{\delta}(\rho_{1}-a+\varepsilon)\bar{\delta}(\rho_{2}-a-\varepsilon)+\bar{\delta}(\rho_{1}-a-\varepsilon)\bar{\delta}(\rho_{2}-a+\varepsilon)\right],\nonumber\\
\bar{\delta}(\rho_{3}-b)\bar{\delta}(\rho_{4}-b)&\rightarrow \frac{1}{2}\left[\bar{\delta}(\rho_{3}-b+\varepsilon)\bar{\delta}(\rho_{4}-b-\varepsilon)+\bar{\delta}(\rho_{3}-b-\varepsilon)\bar{\delta}(\rho_{4}-b+\varepsilon)\right],
\end{align}
where we defined the ratios
\begin{equation}
a \colonequals \frac{(C*A_1B_1)}{(D*A_1B_1)}, \quad\mathrm{and} \quad b \colonequals \frac{(C*A_2B_2)}{(D*A_2B_2)}\, . 
\end{equation} 
Miraculously, this completely removes the $1/0$-like divergences and 
one is left with
\begin{align}\label{loop-integral}
&\langle\big( \phi_{ab}\phi_{cd}\big)^i_{\phantom{i}j}(x_1) \big(\phi_{a'b'}\phi_{c'd'}\big)^k_{\phantom{k}l}(x_2) \rangle_{1-\mathrm{loop}}=\\
&=\frac{16 \delta^{i}_{l}\delta^{k}_{j}g^2_{YM}N^2}{4 (2\pi)^8}   \int\frac{\dd^{4}z}{|x_1-z|^4|z-x_2|^4} \left(\frac{1}{2} \epsilon_{abcd}\epsilon_{a'b'c'd'}-\epsilon_{aba'b'}\epsilon_{c'd'cd}+\epsilon_{abc'd'}\epsilon_{a'b'cd}\right).\nonumber
\end{align}
Note the factor of $16$ in the numerator coming from the four factors of $2$ in the propagator \eqref{eq:treelevelpropagator}. In extracting the anomalous dimensions, two factors of $2/(2\pi)^2$ vanish due to the normalization of the tree-level propagator \eqref{eq:treelevelpropagator}.
Furthermore, we obtain an extra factor of $2\pi^2$ due to the regularization of the UV divergent integral. Finally, we then extract from \eqref{loop-integral} the one-loop dilatation operator in the SO$(6)$ sector
\beq
\Gamma=\frac{g_{YM}^2N}{8\pi^2}\sum_{\ell=1}^L\left(1-P_{\ell,\ell+1}+\frac{1}{2}K_{\ell,\ell+1}\right).
\eeq
Here $P$ is the permutation operator and $K$ the trace operator, see \cite{Minahan:2002ve, Minahan:2010js} for further details.

Let us now conclude with a few words. In this paper we used the twistor space action developed in \cite{Boels:2006ir} to compute an intrinsically off-shell quantity---the planar two-point correlation function of states in the SO$(6)$ sector. In doing so, we encountered naive divergences of the loop {\it integrand}, which result from twistors on the interaction line being forced by the propagators to coincide. Thanks to a symmetric line-splitting procedure, we saw that these divergences are spurious, and we managed to obtain the well-known dilatation operator. An essential observation was that the whole computation employed only a single twistor space diagram, which encourages us to think that the integrable properties of $\mathcal{N}=4$ SYM should become significantly easier to see in twistor space.

We hope to initiate with this note the beginning of a program for understanding the origins of {\it quantum} integrability in $\mathcal{N}=4$ SYM by using the twistor formalism. In addition, while so far the twistor approach has mostly been applied to the calculation of on-shell quantities like amplitudes, we hope that this article  and the forthcoming one \cite{KMSW}  show its usefulness in  investigating off-shell quantities as well. 
In the latter article, we will present in detail further indications of the power of the twistor approach as well as its applications to form factors, see the contemporaneously appearing paper \cite{Wilhelm:2014qua}.

\section*{Acknowledgments}
We are thankful to Rutger Boels, Mathew Bullimore, Burkhard Eden, Jan Fokken, Nils Kanning, Pedro Liendo, Lionel Mason, Christoph Sieg and, especially, Matthias Wilhelm for insightful comments and numerous discussions. We thank the Theory Group at CERN for their hospitality during the crucial stage of preparation of this work. We furthermore gratefully acknowledge support from the Simons Center for Geometry and Physics, Stony Brook University, as well as of the C.N.\ Yang Institute for Theoretical Physics,  where some of the research for this paper was performed. L.K.\ and V.M.\ acknowledge the support of the Marie Curie International Research Staff Exchange Network UNIFY of the European Union's Seventh Framework Programme [FP7-People-2010-IRSES] under grant agreement n°269217, which allowed them to visit Stony Brook University. This research is also supported in part by the SFB 647 \emph{``Raum-Zeit-Materie. Analytische und Geometrische Strukturen''} and the Marie Curie network GATIS (\texttt{\href{http://gatis.desy.eu}{gatis.desy.eu}}) of the European Union’s Seventh Framework Programme FP7/2007-2013/ under REA Grant Agreement No 317089.


\providecommand{\href}[2]{#2}\begingroup\raggedright\endgroup

\end{document}